\documentclass[aps,pre,preprint,groupedaddress]{revtex4-1}
\usepackage[T1]{fontenc}
\usepackage{graphicx}
\usepackage{amsmath}
\usepackage{listings}
\usepackage{url}

\begin{document}

\lstset{language=Python}
\lstset{tabsize=4}              
\lstset{frame=lines}
\lstset{showstringspaces=false}   

\title{Weighted graph algorithms with Python}

\author{A. Kapanowski}

\email[Corresponding author: ]{andrzej.kapanowski@uj.edu.pl}

\author{{\L}. Ga{\l}uszka}

\affiliation{Marian Smoluchowski Institute of Physics, 
Jagiellonian University, ulica {\L}ukasiewicza 11, 30-048 Krak\'{o}w, Poland}

\date{\today}

\begin{abstract}
Python implementation of selected weighted graph 
algorithms is presented. 
The minimal graph interface is defined together with several
classes implementing this interface.
Graph nodes can be any hashable Python objects.
Directed edges are instances of the \lstinline|Edge| class.
Graphs are instances of the \lstinline|Graph| class.
It is based on the adjacency-list representation, but with fast 
lookup of nodes and neighbors (dict-of-dict structure).
Other implementations of this class are also possible.

In this work, many algorithms are implemented using a unified approach. 
There are separate classes and modules devoted to different algorithms.
Three algorithms for finding a minimum spanning tree are implemented:
the  Bor\.{u}vka's algorithm,
the Prim's algorithm (three implementations), and
the Kruskal's algorithm.
Three algorithms for solving the single-source shortest path problem 
are implemented:
the dag shortest path algorithm,
the Bellman-Ford algorithm, and
the Dijkstra's algorithm (two implementations).
Two algorithms for solving all-pairs shortest path problem
are implemented:
the Floyd-Warshall algorithm and
the Johnson's algorithm.

All algorithms were tested by means of the \lstinline|unittest| module,
the Python unit testing framework.
Additional computer experiments were done in order to compare
real and theoretical computational complexity.
The source code is available from the public GitHub repository.
\end{abstract}

\maketitle

\section{Introduction}
\label{sec:intro}

Algorithms are at the heart of computer science.
They are expressed as a finite sequence of operations
with clearly defined input and output.
Algorithms can be expressed by natural languages, pseudocode,
flowcharts or programming languages.
Pseudocode is often used in textbooks and scientific publications
because it is compact and augmented with natural language description.
However, in depth understanding of algorithms is almost
impossible without computer experiments where
algorithms are implemented by means of a programming language.

Our aim is to show that the Python programming language
\cite{python}
can be used to implement algorithms with simplicity and elegance.
On the other hand, the code is almost as readable as pseudocode
because the Python syntax is very clear.
That is why Python has become one of the most popular teaching 
languages in colleges and universities
\cite{2015_Shein}, and it is used in scientific research
\cite{2015_Perkel}.
Python implementation of some algorithms from computational
group theory was shown in Ref. 
\cite{2014_Kapanowski}.
In this paper we are interested in graph theory and
weighted graph algorithms. 
Other algorithms will be discussed elsewhere.
The source code of our programs is available from the public
GitHub repository \cite{graph-dict}. We strongly support
a movement toward open source scientific software
\cite{2015_Hayden}.

Graphs can be used to model many types of relations and processes 
in physical, biological, social, and information systems
\cite{wiki_graph_theory}.
Many practical problems can be represented by graphs
and this can be the first step to finding the solution of the problem.
Sometimes the solution can be known for a long time because
graph theory was born in 1736. In this year Leonard Euler
solved the famous Seven Bridges of K\"{o}nigsberg 
(Kr\'{o}lewiec in Polish) problem.

We examined several Python graph packages
from The Python Package Index \cite{pypi}
in order to checked different approaches.
Some graph libraries written in other programming languages
were checked briefly
\cite{boost},
\cite{boost_graph},
\cite{leda},
\cite{leda_book},
\cite{gtl}.

\begin{itemize}
\item
Package \emph{NetworkX 1.9} by Aric A. Hagberg, Daniel A. Schult,
and Pieter J. Swart
\cite{networkx}.
A library for the creation, manipulation, and study of the structure, 
dynamics, and functions of complex networks.
Basic classes are \lstinline|Graph|, \lstinline|DiGraph|,
\lstinline|MultiGraph|, and \lstinline|MultiDiGraph|.
All graph classes allow any hashable object as a node.
Arbitrary edge attributes can be associated with an edge.
The dictionary of dictionaries data structure is used
to store graphs. 
NetworkX is integrated into Sage.
\cite{sage}.
\item
Package \emph{python-igraph 0.7.0}
\cite{python-igraph}.
\emph{igraph} is the network analysis package with versions 
for R, Python, and C/C++.
\item
Package \emph{python-graph 1.8.2} by Pedro Matiello
\cite{python-graph}.
A library for working with graphs in Python.
Edges are standard Python tuples, weights or labels are kept separately.
There are different classes for directed graphs, undirected graphs,
and hypergraphs.
\item
Package \emph{graph 0.4} by Robert Dick and Kosta Gaitanis
\cite{Dick_Gaitanis}.
Directed and undirected graph data structures and algorithms.
\end{itemize}

None of the available implementations satisfy our needs.
Usually high computational efficiency leads to the unreadable code.
On the other hand, C/C++ syntax or Java syntax are not very close
to the pseudocode used for learning algorithms.
That is why we develop and advocate our approach.
We note that presented Python implementations may be not as fast as
C/C++ or Java counterparts but they scale with the input size
according to the theory.
The presented algorithms are well known and that is why we used
references to many Wikipedia pages.

The paper is organized as follows.
In Section \ref{sec:definitions} basic definitions 
from graph theory are given.
In Section \ref{sec:graph_iface} the graph interface is presented.
In Sections \ref{sec:mst}, \ref{sec:single-source}, and
\ref{sec:all-pairs} the following algorithms are shown:
for finding a minimum spanning tree,
for solving the single-source shortest path problem,
for solving all-pair shortest path problem.
Conclusions are contained in Section \ref{sec:conclusions}.

\section{Definitions}
\label{sec:definitions}

Definitions of graphs vary and that is why we will present
our choice which is very common
\cite{CLRS}.
We will not consider multigraphs with loops and multiple edges.

\subsection{Graphs}

A \emph{(simple) graph} is an ordered pair $G=(V,E)$,
where $V$ is a finite set of nodes (vertices, points)
and $E$ is a finite set of edges (lines, arcs, links).
An edge is an ordered pair of \emph{different} nodes from $V$, $(s,t)$, 
where $s$ is the source node and $t$ is the target node.
This is a \emph{directed edge} and in that case $G$ is 
called a \emph{directed graph}.

An edge can be defined as a 2-element subset of $V$, $\{s,t\}=\{t,s\}$,
and then it is called an \emph{undirected edge}.
The nodes $s$ and $t$ are called the ends of the edge (endpoints)
and they are \emph{adjacent} to one another.
The edge connects or joins the two endpoints.
A graph with undirected edges is called an \emph{undirected graph}.
In our approach, an undirected edge 
corresponds to the set of two directed edges $\{(s,t), (t,s)\}$
and the representative is usually $(s,t)$ with $s<t$.

The \emph{order} of a graph $G=(V,E)$ is the number of nodes $|V|$.
The \emph{degree} of a node in an undirected graph is the number 
of edges that connect to it.

A graph $G'=(V',E')$ is a \emph{subgraph} of a graph $G=(V,E)$
if $V'$ is a subset of $V$ and $E'$ is a subset of $E$.

\subsection{Graphs with weights}

A graph structure can be extended by assigning a number (weight)
$w(s,t)$ to each edge $(s,t)$ of the graph. 
Weights can represent lengths, costs or capacities.
In that case a graph is a \emph{weighted graph}.

\subsection{Paths and cycles}

A \emph{path} $P$ from $s$ to $t$ in a graph $G=(V,E)$ is a sequence 
of nodes from $V$, $(v_0,v_1,\ldots,v_n)$,
where $v_0=s$, $v_n=t$, and $(v_{i-1},v_{i})$
($i=1,\ldots,n$) are edges from $E$.
The length of the path $P$ is $n$.
A \emph{simple path} is a path with distinct nodes.
The \emph{weight (cost or length) of the path} in a weighted graph 
is the sum of the weights of the corresponding edges.

A \emph{cycle} is a path $C$ starting and ending at the same node,
$v_0 = v_n$.
A \emph{simple cycle} is a cycle with no repetitions of nodes allowed, 
other than the repetition of the starting and ending node.

A \emph{directed path (cycle)} is a path (cycle) where 
the corresponding edges are directed.
In our implementation, a path is a list of nodes.
A graph is \emph{connected} if for every pair of nodes $s$ and $t$,
there is a path from $s$ to $t$.

\subsection{Trees}

A \emph{(free) tree} is a connected undirected graph $T$ with no cycles.
A \emph{forest} is a disjoint union of trees.
A \emph{spanning tree} of a connected, undirected graph $G$
is a tree $T$ that includes all nodes of $G$ and is a subgraph of $G$
\cite{wiki_spanning_tree}.
Spanning trees are important because they construct a sparse subgraph
that tells a lot about the original graph.
Also some hard problems can be solved approximately by using spanning 
trees (e.g. traveling salesman problem).

A \emph{rooted tree} is a tree $T$ where one node is designated 
the \emph{root}.
In that case, the edges can be oriented towards or away from the root.
In our implementation, a rooted tree is kept as a dictionary,
where keys are nodes and values are parent nodes.
The parent node of the root is \lstinline|None|.
A forest of rooted trees can be kept in a dictionary
with many roots.

A \emph{shortest-path tree} rooted at node $s$ is a spanning tree $T$ of $G$, 
such that the path distance from root $s$ to any other node $t$ in $T$ 
is the shortest path distance from $s$ to $t$ in $G$.

\section{Interface for graphs}
\label{sec:graph_iface}

According to the definitions from Section \ref{sec:definitions},
graphs are composed of nodes and edges.
In our implementation nodes can be any hashable object that can be sorted.
Usually they are integer or string.

Edges are instances of the \lstinline|Edge| class (\lstinline|edges| module)
and they are directed, hashable, and comparable.
Any edge has the starting node (\lstinline|edge.source|),
the ending node (\lstinline|edge.target|),
and the weight (\lstinline|edge.weight|).
The default weight is one.
The edge with the opposite direction is equal to \lstinline|~edge|.
This is very useful for combinatorial maps used to represent
planar graphs
\cite{wiki_combinatorial_maps}.

Simple graphs (directed and undirected) are instances 
of the \lstinline|Graph| class (\lstinline|graphs| module).
Multigraphs (directed and undirected) are instances of the 
\lstinline|MultiGraph| class (\lstinline|multigraphs| module)
and they will not be discussed here.
Let us show some properties of graphs that are listed
in Table \ref{tab:class_graph}.
There are methods to report some numbers (nodes, edges, degrees).
There are iterators over nodes and edges.
There are also some logical functions.

\begin{lstlisting}
>>> from edges import Edge
>>> from graphs import Graph
>>> G = Graph(n=3, directed=False)
# n is for compatibility with other implementations.
>>> G.is_directed()
False
>>> G.add_edge(Edge('A', 'B', 5))
>>> G.add_edge(Edge('A', 'C', 7))
# Nodes are added by default.
>>> print G.v(), G.e()        # numbers of nodes and edges
3 2
>>> list(G.iternodes())
['A', 'C', 'B']               # random order of nodes
>>> sorted((G.degree(v) for v in G.iternodes()), reverse=True)
[2, 1, 1]                     # the degree sequence
>>> list(v for v in G.iternodes() if G.degree(v) == 1)
['C', 'B']                    # leafs
>>> sum(edge.weight for edge in G.iteredges())
12                            # the graph (tree) weight
# Typical usage of an algorithm Foo from the module foo.
>>> from foo import Foo
>>> algorithm = Foo(G)        # initialization
>>> algorithm.run()           # calculations
>>> print algorithm.result    # results can be more...
\end{lstlisting}

Note that in the case of undirected graphs, 
\lstinline|edge| and \lstinline|~edge| are two representatives
of the same undirected edge. The method \lstinline|iteredges|
returns a representative with the ordering
\lstinline|edge.source < edge.target|.

Graph algorithms are implemented using a unified approach.
There is a separate class for every algorithm.
Different implementations of the same algorithm are grouped
in one module.
In the \lstinline|__init__| method, main variables and
data structures are initialized.
The name space of a class instance is used to access the variables
and that is why interfaces of the auxiliary methods can be very short.

\begin{table}
\caption[Interface for graphs.]{
\label{tab:class_graph}
Interface for graphs; $G$ is a graph, $s$ and $t$ are nodes.
}
\begin{center}
\begin{tabular}{ll}
\hline\hline
Method name & Short description
\\ \hline
\lstinline|Graph(n)| & return an empty undirected graph   \\
\lstinline|Graph(n, directed=True)| & return an empty directed graph   \\
\lstinline|G.is_directed()| & return \lstinline|True| if $G$ is a directed graph \\
\lstinline|G.v()| & return the number of nodes   \\
\lstinline|G.e()| & return the number of edges  \\
\lstinline|G.add_node(s)| & add $s$ to $G$    \\
\lstinline|G.del_node(s)| & remove $s$ form $G$    \\
\lstinline|G.has_node(s)| & return \lstinline|True| if $s$ is in $G$    \\
\lstinline|G.add_edge(edge)| & add a new edge to $G$    \\
\lstinline|G.del_edge(edge)| & remove the edge form $G$    \\
\lstinline|G.has_edge(edge)| & return \lstinline|True| if the edge is in $G$    \\
\lstinline|G.weight(edge)| & return the edge weight or zero    \\
\lstinline|G.iternodes()| & generate nodes on demand  \\
\lstinline|G.iteredges()| & generate edges on demand  \\
\lstinline|G.iteroutedges(s)| & generate outedges on demand  \\
\lstinline|G.iterinedges(s)| & generate inedges on demand  \\
\lstinline|G.degree(s)| & return the degree of $s$ ($G$ undirected)   \\
\lstinline|G.indegree(s)| & return the indegree of $s$    \\
\lstinline|G.outdegree(s)| & return the outdegree of $s$    \\
\hline\hline
\end{tabular}
\end{center}
\end{table}

\section{Minimum spanning tree}
\label{sec:mst}

Let us assume that $G=(V,E)$ is a connected, undirected, weighted graph,
and $T$ is a spanning tree of $G$.
We can assign a weight to the spanning tree $T$ by computing 
the sum of the weights of the edges in $T$.
A \emph{minimum spanning tree (MST)} is a spanning tree with the weight 
less than or equal to the weight of every other spanning tree
\cite{wiki_mst}.
In general, MST is not unique.
There is only one MST if each edge has a distinct weight
(a proof by contradiction).

Minimum spanning trees have many practical applications:
the design of networks, taxonomy, cluster analysis, and others
\cite{wiki_mst}.
We would like to present three classical algorithms for finding MST
that run in polynomial time.

\subsection{Bor\.{u}vka's algorithm}

The Bor\.{u}vka's algorithm works for a connected graph whose edges
have distinct weights what implies the unique MST.
In the beginning, the cheapest edge from each node to another 
in the graph is found, without regard to already added edges.
Then joining these groupings continues in this way until
MST is completed
\cite{wiki_boruvka}.
Components of MST are tracked using a disjoint-set data structure.
The algorithm runs in $O(E \log V)$ time.

Our implementation of the Bor\.{u}vka's algorithm works also for
disconnected graphs. In that case a forest of minimum spanning trees
is created. What is more, repeated edge weights are allowed
because our edge comparison use also nodes, when weights are equal.

\begin{lstlisting}
from edges import Edge
from unionfind import UnionFind

class BoruvkaMST:
    """Boruvka's algorithm for finding MST."""

    def __init__(self, graph):
        """The algorithm initialization."""
        self.graph = graph
        self.mst = graph.__class__(graph.v()) # MST as a graph
        self.uf = UnionFind()

    def run(self):
        """Executable pseudocode."""
        for node in self.graph.iternodes():
            self.uf.create(node)
        forest = set(node for node in self.graph.iternodes())
        dummy_edge = Edge(None, None, float("inf"))
        new_len = len(forest)
        old_len = new_len + 1
        while old_len > new_len:
            old_len = new_len
            min_edges = dict(((node, dummy_edge)
                for node in forest))
            # Finding the cheapest edges.
            for edge in self.graph.iteredges(): # O(E) time
                source = self.uf.find(edge.source)
                target = self.uf.find(edge.target)
                if source != target:   # different components
                    if edge < min_edges[source]:
                        min_edges[source] = edge
                    if edge < min_edges[target]:
                        min_edges[target] = edge
            # Connecting components, total time is O(V).
            forest = set()
            for edge in min_edges.itervalues():
                if edge is dummy_edge: # a disconnected graph
                    continue
                source = self.uf.find(edge.source)
                target = self.uf.find(edge.target)
                if source != target:   # different components
                    self.uf.union(source, target)
                    forest.add(source)
                    self.mst.add_edge(edge)
            # Remove duplicates, total time is O(V).
            forest = set(self.uf.find(node) for node in forest)
            new_len = len(forest)
            if new_len == 1:   # a connected graph
                break
\end{lstlisting}

We note that the Bor\.{u}vka's algorithm is well suited
for parallel computation.

\subsection{Prim's algorithm}

The Prim's algorithm has the property that the edges in growing $T$ 
always form a single tree.
The weights from $G$ can be negative
\cite{wiki_prim}.
We begin with some node $s$ from $G$ and $s$ is added to the empty $T$.
Then, in each iteration, we choose a minimum-weight edge $(s,t)$,
joining $s$ inside $T$ to $t$ outside $T$.
Then the minimum-weight edge is added to $T$.
This process is repeated until MST is formed.

The performance of Prim's algorithm depends on how the priority 
queue is implemented.
In the case of the first implementation a binary heap is used
and it takes $O(E \log V)$ time.

\begin{lstlisting}
from edges import Edge
from Queue import PriorityQueue

class PrimMST:
    """Prim's algorithm for finding MST."""

    def __init__(self, graph):
        """The algorithm initialization."""
        self.graph = graph
        self.distance = dict((node, float("inf")) 
            for node in self.graph.iternodes())
        self.parent = dict((node, None) 
            for node in self.graph.iternodes()) # MST as a dict
        self.in_queue = dict((node, True) 
            for node in self.graph.iternodes())
        self.pq = PriorityQueue()

    def run(self, source=None):
        """Executable pseudocode."""
        if source is None:   # get first random node
            source = self.graph.iternodes().next()
        self.source = source
        self.distance[source] = 0
        for node in self.graph.iternodes():
            self.pq.put((self.distance[node], node))
        while not self.pq.empty():
            _, node = self.pq.get()
            if self.in_queue[node]:
                self.in_queue[node] = False
            else:
                continue
            for edge in self.graph.iteroutedges(node):
                if (self.in_queue[edge.target] and 
                    edge.weight < self.distance[edge.target]):
                        self.distance[edge.target] = edge.weight
                        self.parent[edge.target] = edge.source
                        self.pq.put((edge.weight, edge.target))
\end{lstlisting}

The second implementation is better for dense graphs ($|E| \sim |V|^2$),
where the adjacency-matrix representation of graphs is often used.
For-loop is executed $|V|$ times, finding the minimum takes $O(V)$.
Therefore, the total time is $O(V^2)$.

\begin{lstlisting}
class PrimMatrixMST:
    """Prim's algorithm for finding MST in O(V**2) time."""

    def __init__(self, graph):
        """The algorithm initialization."""
        self.graph = graph
        self.distance = dict((node, float("inf")) 
            for node in self.graph.iternodes())
        self.parent = dict((node, None) 
            for node in self.graph.iternodes()) # MST as a dict
        self.in_queue = dict((node, True) 
            for node in self.graph.iternodes())

    def run(self, source=None):
        """Executable pseudocode."""
        if source is None:   # get first random node
            source = self.graph.iternodes().next()
        self.source = source
        self.distance[source] = 0
        for step in xrange(self.graph.v()): # |V| times
            # Find min node in the graph, O(V) time.
            node = min((node for node in self.graph.iternodes() 
                if self.in_queue[node]), key=self.distance.get)
            self.in_queue[node] = False
            for edge in self.graph.iteroutedges(node): # O(V) time
                if (self.in_queue[edge.target] and
                    edge.weight < self.distance[edge.target]):
                        self.distance[edge.target] = edge.weight
                        self.parent[edge.target] = edge.source
\end{lstlisting}

\subsection{Kruskal's algorithm}

The Kruskal's algorithm builds the MST in forest
\cite{1956_Kruskal},
\cite{wiki_kruskal}.
In the beginning, each node is in its own tree in forest.
Then all edges are scanned in increasing weight order.
If an edge connects two different trees, then the edge is added 
to the MST and the trees are merged.
If an edge connects two nodes in the same tree, 
then the edge is discarded.

The presented implementation uses a priority queue
in order to sort edges by weights.
Components of the MST are tracked using a disjoint-set data 
structure (the \lstinline|UnionFind| class).
The total time is $O(E \log V)$.

\begin{lstlisting}
from unionfind import UnionFind
from Queue import PriorityQueue

class KruskalMST:
    """Kruskal's algorithm for finding MST."""

    def __init__(self, graph):
        """The algorithm initialization."""
        self.graph = graph
        self.mst = graph.__class__(graph.v()) # MST as a graph
        self.uf = UnionFind()
        self.pq = PriorityQueue()

    def run(self):
        """Executable pseudocode."""
        for node in self.graph.iternodes(): # O(V) time
            self.uf.create(node)
        for edge in self.graph.iteredges(): # O(E*log(V)) time
            self.pq.put((edge.weight, edge))
        while not self.pq.empty():     # |E| steps
            _, edge = self.pq.get()  # O(log(V)) time
            if (self.uf.find(edge.source) != 
                self.uf.find(edge.target)):
                    self.uf.union(edge.source, edge.target)
                    self.mst.add_edge(edge)
\end{lstlisting}

The Kruskal's algorithm is regarded as best when the edges can
be sorted fast or are already sorted.

\section{Single-source shortest path problem}
\label{sec:single-source}

Let us assume that $G=(V,E)$ is a weighted directed graph.
The \emph{shortest path problem} is the problem of finding a path 
between two nodes in $G$ such that the path weight is minimized
\cite{wiki_shortest_path}.
The negative edge weights are allowed but the negative weight cycles
are forbidden (in that case there is no shortest path).

The Dijkstra's algorithm and the Bellman-Ford are based 
on the principle of \emph{relaxation}.
Approximate distances (overestimates) are gradually replaced 
by more accurate values until the correct distances are reached.
However, the details of the relaxation process differ.
Many additional algorithms may be found in Ref.
\cite{1996_Cherkassky}.

\subsection{Bellman-Ford algorithm}

The Bellman-Ford algorithm computes shortest paths from a single 
source to all of the other nodes in a weighted graph $G$
in which some of the edge weights are negative 
\cite{wiki_bellman_ford}.
The algorithm relaxes all the edges and it does this $|V|-1$ times.
At the last stage, negative cycles detection is performed and
their existence is reported.
The algorithm runs in $O(V \cdot E)$ time.

\begin{lstlisting}
class BellmanFord:
    """The Bellman-Ford algorithm for the shortest path problem."""

    def __init__(self, graph):
        """The algorithm initialization."""
        if not graph.is_directed():
            raise ValueError("graph is not directed")
        self.graph = graph
        self.distance = dict(((node, float("inf"))
            for node in self.graph.iternodes()))
        # Shortest path tree as a dictionary.
        self.parent = dict(((node, None)
            for node in self.graph.iternodes()))

    def run(self, source):
        """Executable pseudocode."""
        self.source = source
        self.distance[source] = 0
        for step in xrange(self.graph.v()-1):   # |V|-1 times
            for edge in self.graph.iteredges():   # O(E) time
                self._relax(edge)
        # Check for negative cycles.
        for edge in self.graph.iteredges():   # O(E) time
            if (self.distance[edge.target] > self.distance[edge.source]
                + edge.weight):
                    raise ValueError("negative cycle")

    def _relax(self, edge):
        """Edge relaxation."""
        alt = self.distance[edge.source] + edge.weight
        if self.distance[edge.target] > alt:
            self.distance[edge.target] = alt
            self.parent[edge.target] = edge.source
            return True
        return False

    def path(self, target):
        """Construct a path from source to target."""
        if self.source == target:
            return [self.source]
        elif self.parent[target] is None:
            raise ValueError("no path to target")
        else:
            return self.path(self.parent[target]) + [target]
\end{lstlisting}

\subsection{Dijkstra's algorithm}

The Dijkstra's algorithm solves the single-source shortest path 
problem for a graph $G$ with non-negative edge weights, 
producing a shortest path tree $T$
\cite{1959_Dijkstra},
\cite{wiki_dijkstra}. 
It is a greedy algorithm that starts at the source node,
then it grows $T$ and spans all nodes reachable from the source.
Nodes are added to $T$ in order of distance.
The relaxation process is performed on outgoing edges of
minimum-weight nodes.
The total time is $O(E \log V)$.

\begin{lstlisting}
from Queue import PriorityQueue

class Dijkstra:
    """Dijkstra's algorithm for the shortest path problem."""

    def __init__(self, graph):
        """The algorithm initialization."""
        if not graph.is_directed():
            raise ValueError("graph is not directed")
        self.graph = graph
        self.distance = dict((node, float("inf")) 
            for node in self.graph.iternodes())
        self.parent = dict((node, None) 
            for node in self.graph.iternodes())
        self.in_queue = dict((node, True) 
            for node in self.graph.iternodes())
        self.pq = PriorityQueue()

    def run(self, source):
        """Executable pseudocode."""
        self.source = source
        self.distance[source] = 0
        for node in self.graph.iternodes():
            self.pq.put((self.distance[node], node))
        while not self.pq.empty():
            _, node = self.pq.get()
            if self.in_queue[node]:
                self.in_queue[node] = False
            else:
                continue
            for edge in self.graph.iteroutedges(node):
                if self.in_queue[edge.target] and self._relax(edge):
                    self.pq.put((self.distance[edge.target],
                        edge.target))
\end{lstlisting}

The second implementation is better for dense graphs
with the adjacency-matrix representation, the total time is $O(V^2)$.

\begin{lstlisting}
class DijkstraMatrix:
    """Dijkstra's algorithm with O(V**2) time."""

    def __init__(self, graph):
        """The algorithm initialization."""
        if not graph.is_directed():
            raise ValueError("graph is not directed")
        self.graph = graph
        self.distance = dict((node, float("inf"))
            for node in self.graph.iternodes())
        self.parent = dict((node, None)
            for node in self.graph.iternodes())
        self.in_queue = dict((node, True)
            for node in self.graph.iternodes())

    def run(self, source):
        """Executable pseudocode."""
        self.source = source
        self.distance[source] = 0
        for step in xrange(self.graph.v()):   # |V| times
            # Find min node, O(V) time.
            node = min((node for node in self.graph.iternodes() 
                if self.in_queue[node]), key=self.distance.get)
            self.in_queue[node] = False
            for edge in self.graph.iteroutedges(node): # O(V) time
                if self.in_queue[edge.target]:
                    self._relax(edge)
\end{lstlisting}

Le us note that a time of $O(E+V \log V)$ is best possible
for Dijkstra's algorithm, if edge weights are real numbers
and only binary comparisons are used.
This bond is attainable using Fibonacci heaps, relaxed heaps or Vheaps
\cite{1990_Ahuja}.
Researchers are also working on adaptive algorithms which profit
from graph easiness (small density, not many cycles).

\subsection{DAG Shortest Path}

Shortest paths in DAG are always defined because there are
no cycles. The algorithm runs in $O(V+E)$ time because
topological sorting of graph nodes is conducted
\cite{CLRS}.

\begin{lstlisting}
from topsort import TopologicalSort

class DAGShortestPath:
    """The shortest path problem for DAG."""

    def __init__(self, graph):
        """The algorithm initialization."""
        if not graph.is_directed():
            raise ValueError("graph is not directed")
        self.graph = graph
        self.distance = dict((node, float("inf"))
            for node in self.graph.iternodes())
        self.parent = dict((node, None)
            for node in self.graph.iternodes())

    def run(self, source):
        """Executable pseudocode."""
        self.source = source
        self.distance[source] = 0
        algorithm = TopologicalSort(self.graph)
        algorithm.run()
        for source in algorithm.sorted_nodes:
            for edge in self.graph.iteroutedges(source):
                self._relax(edge)
\end{lstlisting}

\section{All-pairs shortest path problem}
\label{sec:all-pairs}

Let us assume that $G=(V,E)$ is a weighted directed graph.
The \emph{all-pairs shortest path problem} is the problem of finding 
shortest paths between every pair of nodes
\cite{wiki_shortest_path}.
Two algorithms solving this problem are shown:
the Floyd-Warshall algorithm and
the Johnson's algorithm.
There is the third algorithm in our repository, which is based
on matrix multiplication (the \lstinline|allpairs.py| module)
\cite{CLRS}.
A basic version has a running time of $O(V^4)$
(the \lstinline|SlowAllPairs| class), 
but it is improved to $O(V^3 \log V)$
(the \lstinline|FasterAllPairs| class).

\subsection{Floyd-Warshall algorithm}

The Floyd-Warshall algorithm computes shortest paths in a weighted graph 
with positive or negative edge weights, but with no negative cycles
\cite{wiki_floyd_warshal}.
The algorithm uses a method of dynamic programming.
The shortest path from $s$ to $t$ without intermediate nodes
has the length $w(s,t)$.
The shortest paths estimates are incrementally improved
using a growing set of intermediate nodes.
$O(V^3)$ comparisons are needed to solve the problem.
Three nested for-loops are the heart of the algorithm.

Our implementation allows the reconstruction of the path
between any two endpoint nodes.
The shortest-path tree for each node is calculated
and $O(V^2)$ memory is used.
The diagonal of the path matrix is inspected and 
the presence of negative numbers indicates negative cycles.

\begin{lstlisting}
class FloydWarshallPaths:
    """The Floyd-Warshall algorithm with path reconstruction."""

    def __init__(self, graph):
        """The algorithm initialization."""
        if not graph.is_directed():
            raise ValueError("graph is not directed")
        self.graph = graph
        self.distance = dict()
        self.parent = dict()
        for source in self.graph.iternodes():
            self.distance[source] = dict()
            self.parent[source] = dict()
            for target in self.graph.iternodes():
                self.distance[source][target] = float("inf")
                self.parent[source][target] = None
            self.distance[source][source] = 0
        for edge in self.graph.iteredges():
            self.distance[edge.source][edge.target] = edge.weight
            self.parent[edge.source][edge.target] = edge.source

    def run(self):
        """Executable pseudocode."""
        for node in self.graph.iternodes():
            for source in self.graph.iternodes():
                for target in self.graph.iternodes():
                    alt = self.distance[source][node] + \
                        self.distance[node][target]
                    if alt < self.distance[source][target]:
                        self.distance[source][target] = alt
                        self.parent[source][target] = \
                            self.parent[node][target]
        if any(self.distance[node][node] < 0
        for node in self.graph.iternodes()):
            raise ValueError("negative cycle detected")

    def path(self, source, target):
        """Path reconstruction."""
        if source == target:
            return [source]
        elif self.parent[source][target] is None:
            raise ValueError("no path to target")
        else:
            return self.path(source, self.parent[target]) + [target]
\end{lstlisting}

\subsection{Johnson's algorithm}

The Johnson's algorithm finds the shortest paths between all pairs 
of nodes in a sparse directed graph.
The algorithm uses the technique of \emph{reweighting}.
It works by using the Bellman-Ford algorithm to compute a transformation 
of the input graph that removes all negative weights, allowing Dijkstra's 
algorithm to be used on the transformed graph
\cite{wiki_johnson}.
The time complexity of our implementation is $O(V E \log V)$,
because the binary min-heap is used in the Dijkstra's algorithm.
It is asymptotically faster than the Floyd-Warshall algorithm 
if the graph is sparse.

\begin{lstlisting}
from edges import Edge
from bellmanford import BellmanFord
from dijkstra import Dijkstra

class Johnson:
    """The Johnson algorithm for the shortest path problem."""

    def __init__(self, graph):
        """The algorithm initialization."""
        if not graph.is_directed():
            raise ValueError("graph is not directed")
        self.graph = graph

    def run(self):
        """Executable pseudocode."""
        self.new_graph = self.graph.__class__(
            self.graph.v() + 1, directed=True)
        for node in self.graph.iternodes():   # O(V) time
            self.new_graph.add_node(node)
        for edge in self.graph.iteredges():   # O(E) time
            self.new_graph.add_edge(edge)
        self.new_node = self.graph.v()
        self.new_graph.add_node(self.new_node)
        for node in self.graph.iternodes():   # O(V) time
            self.new_graph.add_edge(Edge(self.new_node, node, 0))
        self.bf = BellmanFord(self.new_graph)
        # If this step detects a negative cycle,
        # the algorithm is terminated.
        self.bf.run(self.new_node)   # O(V*E) time
        # Edges are reweighted.
        for edge in list(self.new_graph.iteredges()):   # O(E) time
            edge.weight = (edge.weight 
                + self.bf.distance[edge.source]
                - self.bf.distance[edge.target])
            self.new_graph.del_edge(edge)
            self.new_graph.add_edge(edge)
        # Remove new_node with edges.
        self.new_graph.del_node(self.new_node)
        # Weights are now modified!
        self.distance = dict()
        for source in self.graph.iternodes():
            self.distance[source] = dict()
            algorithm = Dijkstra(self.new_graph)
            algorithm.run(source)
            for target in self.graph.iternodes():
                self.distance[source][target] = (
                    algorithm.distance[target]
                    - self.bf.distance[source] 
                    + self.bf.distance[target])
\end{lstlisting}

\section{Conclusions}
\label{sec:conclusions}

In this paper, we presented Python implementation of several
weighted graph algorithms. The algorithms are represented 
by classes where graph objects are processed via proposed graph
interface. The presented implementation is unique in several ways.
The source code is readable like a pseudocode from textbooks or 
scientific articles. On the other hand, the code can be executed
with efficiency established by the corresponding theory.
Python's class mechanism adds classes with minimum of new
syntax and it is easy to create desired data structures
(e.g., an edge, a graph, a union-find data structure)
or to use objects from standard modules (e.g., queues, stacks).

The source code is available from the public GitHub repository
\cite{graph-dict}. It can be used in education, scientific research,
or as a starting point for implementations in other programming
languages. The number of available algorithms is growing,
let us list some of them:

\begin{itemize}
\item
Graph traversal (breadth-first search, depth-first search)
\item
Connectivity (connected components, strongly connected components)
\item
Accessibility (transitive closure)
\item
Topological sorting
\item
Cycle detection
\item
Testing bipartiteness
\item
Minimum spanning tree
\item
Matching (Augmenting path algorithm, Hopcroft-Karp algorithm)
\item
Shortest path search
\item
Maximum flow (Ford-Fulkerson algorithm, Edmonds-Karp algorithm)
\item
Graph generators
\end{itemize}






\end{document}